\documentclass[%
 reprint,
 amsmath,amssymb,
 aps,
]{revtex4-1}

\usepackage{graphicx}% Include figure files
\usepackage{dcolumn}% Align table columns on decimal point
\usepackage{bm}% bold math
\usepackage[version=3]{mhchem} % Formula subscripts using \ce{}
\usepackage[T1]{fontenc}       % Use modern font encodings
\usepackage[colorlinks]{hyperref}
\usepackage{color,soul} 
\usepackage{textcomp}
\usepackage{soul}
\usepackage[pagewise]{lineno}
\usepackage{float}
\usepackage{lineno}
\usepackage{amssymb}

\usepackage{amsmath,amssymb}
\usepackage{graphicx}% Include figure files
\usepackage{dcolumn}% Align table columns on decimal point
\usepackage{soul}
\usepackage{amssymb}
\usepackage[pagewise]{lineno}
\usepackage{float}
\usepackage{xr}
\begin{document}

\title{Probing Large Deformation and Fracture Behavior of Physically Assembled Gel System by Varying Polymer Concentration$^\dagger$}% Force line breaks with \\
\thanks{Supporting information available}%

\author{Satish Mishra}
\affiliation{%
	Dave C. Swalm School of Chemical Engineering, Mississippi State University, MS State, MS 39762.
}%
\author{Rosa Maria Badani Prado}%
\affiliation{%
 Dave C. Swalm School of Chemical Engineering, Mississippi State University, MS State, MS 39762.
}%
\author{Santanu Kundu}%
\email{santanukundu@che.msstate.edu}
\affiliation{%
	Dave C. Swalm School of Chemical Engineering, Mississippi State University, MS State, MS 39762.
}%

\date{\today}

\begin{abstract}
	
Physically assembled gels have promising applications in many fields because of their tunable mechanical properties. Here, we report the mechanical properties as a function of polymer volume fraction ($\phi$) for a physical gel system consists of poly(styrene)-poly(isoprene)-poly(styrene) [PS-PI-PS] in mineral oil. The PI-block molecular weight is higher than the entanglement molecular weight, which leads to the entanglement of PI-blocks at higher $\phi$. The micellar microstructure for all gels results in a similar stress relaxation mechanism, as captured by the superposition of stress-relaxation results. Tensile testing experiments reveal a strain-rate dependence mechanical response for the entangled gels. To capture the critical energy release rate ($\Gamma_0$) over a range of $\phi$, both cavitation rheology and fracture experiments were performed and we obtain $\Gamma_0\sim\phi^{2.0}$. The gel moduli scale with the volume fraction as $\phi^{2.39}$, where the exponent is likely dictated by the change in loop-to-bridge fraction with increasing $\phi$.

\end{abstract}

\pacs{Valid PACS appear here}
\maketitle

\section{Introduction}
Physically assembled gels are used in many applications including in prosthetic, protective clothing, tissue surrogates, and consumer orthotics products such as shoe inserts.\cite{Chen2007Patent,Ford2010} Out of many varieties, one of the common physical gels consists of ABA triblock copolymers in B-block selective solvents. Like any materials, the performance of these gels depends on their microstructure. \cite{Laurer1999,Vega2001,Seitz2007,Seitz2009,Erk2010,Erk2012,Erk2012a,Hashemnejad2015,Zabet2015,Zabet2017} The microstructure of these gels originates from a strong temperature-dependent solubility difference of A-block (endblock) and B-block (midblock) in the chosen solvent. As the temperature is decreased from an elevated temperature to below the gelation temperature, a viscous liquid transforms into a gel-like material, as the poorly-soluble A-blocks collapse to form aggregates (crosslinks), and the soluble B-blocks bridge these aggregates and act as load-bearing chains.\cite{Laurer1999,Vega2001,Seitz2007,Seitz2009,Erk2010,Erk2012,Erk2012a,Hashemnejad2015,Zabet2015,Peters2016,Zabet2017}

In conventional chemically crosslinked gels, the change in microstructure and the corresponding mechanical properties with respect to the polymer volume fraction are reasonably well-understood.\cite{rubinstein2003polymer}  With increasing concentration, the polymer chains form entanglements leading to the rate-dependent mechanical properties.\cite{rubinstein2003polymer} However, such a conclusion cannot be drawn directly for the ABA type physical gels considered here, as the block length and polymer volume fraction dictate the microstructure and the resultant mechanical properties in a complex, interdependent manner. For the lower polymer concentration and shorter midblock length, the chains (bridges) connecting the aggregates can be in the stretched-state, and if the corresponding entropic penalty is high, they prefer to form loops with two end blocks located in the same aggregate. At higher polymer volume fraction and longer midblock length, the midblocks can form entanglement. Also, the length of endblocks affects the relaxation behavior by dictating the size and density of aggregates. The exchange of endblocks among the aggregates and endblock pullout form an aggregate subjected to mechanical deformation are important considerations in dictating the mechanical behavior. A number of studies have been conducted to link the gel mechanical properties as a function of gel microstructure, but a fundamental understanding regarding the structure-property relationship is still incomplete.\cite{Laurer1999,Vega2001,Seitz2007,Hashemnejad2015,Zabet2015,Peters2016,Zabet2017}

In our previous study, we have investigated triblock copolymer gels with varying midblock length while maintaining a similar length of endblocks.\cite{Mishra2019b} In this study, we elucidate the effect of polymer concentration on gel mechanical behavior. The midblock length is selected in such a way that it forms entanglement at higher concentration.  For these gels, by changing the polymer concentration, the gel modulus can be varied significantly, which makes it difficult to conduct mechanical characterization using a single tool. Over the last few decades, numerous techniques like shear-rheology,\cite{Erk2010,Zabet2015,Zabet2017} puncture mechanics,\cite{Fakhouri2015,Rattan2018} cavitation rheology,\cite{Zimberlin2007,Kundu2009,Mishra2018,Bentz2018,Yang2019} contact mechanics,\cite{Kundu} tensile tests,\cite{Mrozek2015} fracture with a predefined crack,\cite{Baumberger2006,Seitz2007,Mishra2018a,Mishra2019b} and peel-tests\cite{Tanaka2000} have been implemented to probe these materials, however, each of these techniques is applicable only over a range of modulus and provides very specific information. In our approach, we have used a combination of techniques to investigate the gel mechanical properties over a wide range of polymer volume fraction ($\phi$).

We have considered a gel-system consists of poly(styrene)-poly(isoprene)-poly(styrene) [PS-PI-PS] triblock copolymer (Kraton D1164) dissolved in mineral oil, a midblock selective solvent. The $\phi$ is varied from 0.044 to 0.181, which corresponds to the PS and PI weight fractions in the range of 0.015-0.058, and 0.035-0.142, respectively (Table~\ref{TableConcGelDetails}). Low-strain behavior of these gels has been probed by shear-rheometry while the responses at large deformation for high $\phi$ values have been investigated using a custom-built tensile setup. The critical energy release rate has been determined from cavitation rheology (low $\phi$) and fractures experiments with a predefined crack (high $\phi$). We aim to determine the effect of polymer volume fraction on elastic modulus, large strain deformation subjected to different strain rate, and fracture energy-release rate. 

\begin{table}
	\begin{center}
		\caption{Gel nomenclature, polymer weight fraction, polymer volume fraction, and PS and PI weight fractions in respective gels.}
		\label{TableConcGelDetails}
		\begin{tabular}{lllll}
			\hline
			\multicolumn{1}{p{1.7cm}}{\centering  $\phi$}  &
			\multicolumn{1}{p{1.0cm}}{\centering \% wt}  &   
			\multicolumn{1}{p{1.8cm}}{\centering PS (wt\%) } &
			\multicolumn{1}{p{1.8cm}}{\centering PI (wt\%) } \\

			\hline
			\multicolumn{1}{p{1.7cm}}{\centering 0.044}  &
			\multicolumn{1}{p{1.0cm}}{\centering 5}  &   
			\multicolumn{1}{p{1.8cm}}{\centering 0.015} &
			\multicolumn{1}{p{1.8cm}}{\centering 0.035} \\

			\multicolumn{1}{p{1.7cm}}{\centering 0.067}  &
			\multicolumn{1}{p{1.0cm}}{\centering 7.5}  &   
			\multicolumn{1}{p{1.8cm}}{\centering 0.218} &
			\multicolumn{1}{p{1.8cm}}{\centering 0.053} \\
			
			\multicolumn{1}{p{1.7cm}}{\centering 0.089}  &
			\multicolumn{1}{p{1.0cm}}{\centering 10}  &   
			\multicolumn{1}{p{1.8cm}}{\centering 0.029} &
			\multicolumn{1}{p{1.8cm}}{\centering 0.071 } \\
			
			\multicolumn{1}{p{1.7cm}}{\centering 0.135}  &
			\multicolumn{1}{p{1.0cm}}{\centering 15}  &   
			\multicolumn{1}{p{1.8cm}}{\centering 0.044} &
			\multicolumn{1}{p{1.8cm}}{\centering 0.106} \\
			
			\multicolumn{1}{p{1.7cm}}{\centering 0.181}  &
			\multicolumn{1}{p{1.0cm}}{\centering 20}  &   
			\multicolumn{1}{p{1.8cm}}{\centering 0.058} &
			\multicolumn{1}{p{1.8cm}}{\centering 0.142} \\

			\hline
		\end{tabular}
	\end{center}
\end{table}

%The polymer has an overall number average and weight average molecular weight of  $M_n$=112~kg/mol and $M_w$=125~kg/mol, respectively, with the PI content of 71\%.\cite{Mishra2018a} 
\section{Materials and Methods}

\textbf{Materials.} Kraton D1164 polymer (kindly provided by Kraton Inc.) and Klearol white mineral oil (kindly provided by Sonneborn Inc.) were used here for gel preparation \cite{Mishra2018a,Mishra2019b}. D1164 is a triblock copolymer consist of  poly(styrene)-poly(isoprene)-poly(styrene) [PS- PI-PI]. Based on the manufacturer datasheet, the average PS content of the polymer is 29\%~(w/w) and the PI content is 71\%~(w/w). The molecular weight of the D1164 was determined by using Agilent GPC at 135~$^\circ$C using 1,2,4-trichlorobenzene as a solvent.  Based on the polystyrene standards, the overall number average molecular weight ($M_n$) and weight average molecular weight ($M_w$) were determined as  $\approx$112 kg/mol and $\approx$125 kg/mol, respectively. Based on the wt\% information given by the manufacturer, we estimate the PS-block molecular weight as 16.2~kg/mol and PI-block weight as 79.4~kg/mol.

\textbf{Sample preparation.} The desired amount of 20\% (w/w) polymer was dissolved in mineral oil using a magnetic stirrer (320~rpm) at an elevated temperature (100-140~$^\circ$C) \cite{Mishra2018a,Mishra2019b}. The sample was then placed in the convective oven at 110~$^\circ$C to allow the bubble to escape and to obtain a clear solution. As shown in Table~\ref{TableConcGelDetails}, five different volume fractions, 0.044, 0.067, 0.089, 0.135, and 0.181 were considered. Table~\ref{TableConcGelDetails} also displays the polymer weight fraction, and the PS and PI weight fractions in each gel.

\textbf{Rheology experiments.} For the rheology experiments, the polymer solution was poured in a rectangular mold to obtain a gel sheet of 2~mm thickness. Small samples (30~mm$\times$30~mm) were then cut from this sheet for the rheology experiments. A TA Instruments Discovery HR-2 hybrid rheometer equipped with a Peltier plate was used. Experiments were conducted using 25~mm diameter parallel plate geometry. To avoid the sample slippage, a 240 grit adhesive-backed silicon carbide sandpaper (Allied High Tech Products Inc.) was attached to the top and bottom plates.

\textbf{Scattering experiments.} Small-angle X-ray scattering (SAXS) experiments were conducted on the gels with $\phi$=0.089, 0.135, and 0.181.  All experiments were conducted at room temperature (22~$^\circ$C) with a sample-to-detector distance of 2.5 m using the Xeuss 2.0 beamline. The data collection was performed using a Pilatus 1M detector. Igor 8 Nika package was used for data reduction from 2D scattering profile to 1D intensity ($I(q)$) versus scattering vector ($q$) plots.\cite{Ilavsky2012} Further, a custom-built NCNR macro was used to fit the data with polydispersed core hard-sphere.\cite{Kline2006}

\textbf{Cavitation experiments.} Cavitation rheology (CR) experiments were performed at room temperature using a 10~mL syringe. Most of our experiments were conducted with a needle having an inner radius of $r_{in}\approx$0.317~mm and an outer radius of $r_{out}$=0.454~mm. The needle was inserted $\approx$15~mm inside the gel and was then retracted 3.45~mm to create a cylindrical cavity. Air was injected at the rate of 5~mL/min in order to expand the defect. In many occasions, the air was noticed to escape from the needle-gel interface rather than expanding the cavity. To circumvent that we have applied acrylic paint on the needle to strengthen the adhesion between the needle wall and gel. However, higher pressure necessary for high polymer volume fraction still caused leaking, therefore, the CR experiments could not be performed on $\phi$=0.135 and 0.181 gels. Therefore, CR experiments were only conducted on $\phi$=0.044, 0.067, and 0.089 mostly using the needle with $r_{out}$=0.454~mm. For establishing the $P-r_{out}$ relationship, needles with $r_{out}\approx$ 0.455, 0.635, 0.8, and 0.915~mm were used. At least three CR experiments were conducted for each case and an average is reported here.

\textbf{Tensile experiments.} The tensile experiments were performed at room temperature using a custom-built set-up controlled by a NI LabVIEW program. For these experiments, dog bone-shaped samples were prepared by casting the hot polymer solution in a dogbone mold \cite{Mrozek2015}. The gauge region of these samples had a height of 4.2 mm, breadth of 4.2 mm, and a thickness of 9.5 mm, respectively. The samples were supported on the experimental set-up by using two pairs of supporting pins. The block attached to the top supporting pin was moved to apply the stretch through a moving stage (PI M414.1PD). Tensile tests were performed on $\phi$=0.089, 0.135, and 0.181 using three different stretch-rates viz. $\dot{\lambda}$ = 0.0048, 0.048 and 0.48~s$^{-1}$, respectively \cite{Mishra2018a,Mishra2019b}. A monochrome camera (Grasshopper3, Point Gray Research Inc.) at 17 frames per second was used to capture images during the experiments. The gauge region of samples was marked with three lines for the purpose of stretch an stretch rate measurements through an image processing program in MATLAB, while the resultant force was measured based on the bending of a cantilever attached to the top supporting block.\cite{Mishra2018a,Mishra2019b}

\textbf{Fracture experiments with a predefined crack.} Fracture experiments were performed using a custom-built set-up.\cite{Baumberger2006} Gel samples were prepared by casting the hot polymer solution into a mold with a width of 80~mm, thickness of 3-4~mm, and a height of 35~mm. In order to observe the fracture at a known position, a sharp cut of length 5~mm was introduced in the middle of the sample. The samples were stretched continuously with a stretch rate of 0.01~s$^{-1}$ until the sample split into two halves. Data post-processing and image processing MATLAB program was implemented for estimating crack- tip velocity.\cite{Mishra2018a,Mishra2019b} Further details regarding our set-up and experimental protocol can be found in our previous work.\cite{Mishra2018a, Mishra2019b}

\section{Results and Discussion}

\begin{figure}
	\begin{center}
		\includegraphics[width=3in]{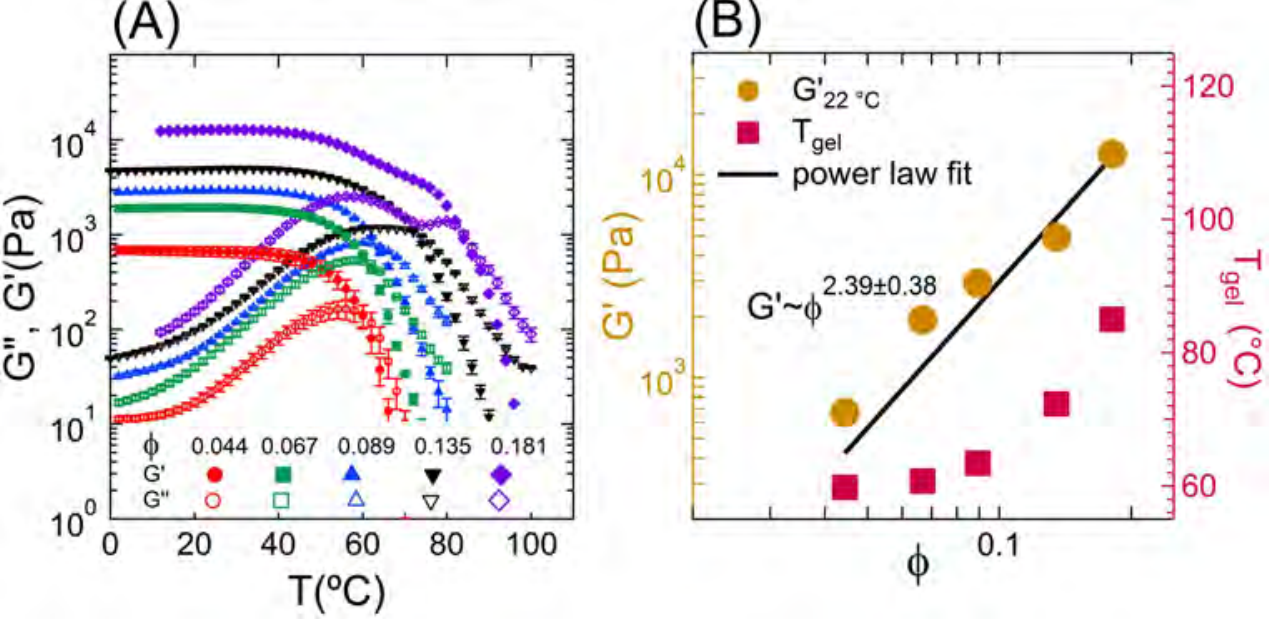}
		%\vspace{-3em}
		\caption{(A) Evolution of storage ($G^{\prime}$) and loss moduli ($G^{\prime\prime}$) as a function of temperature measured using oscillation frequency ($\omega$) of 1~rad/s and amplitude ($\gamma_0$) of 0.01 for the polymer volume fractions ($\phi$) of 0.044, 0.067, 0.089, 0.135, and 0.181, respectively. (B) $G^{\prime}$ at room temperature and $T_{gel}$ as a function of $\phi$.  The solid line indicates a power law fit. Error bars represent one standard deviation obtained from three repeats. }
		\label{PicRheology2}
	\end{center}
\end{figure}

The gelation process has been captured by tracking the storage ($G^{\prime}$) and loss moduli ($G^{\prime\prime}$) as a function of temperature ($T$). The samples behave like a viscous liquid ($G^{\prime\prime}$ > $G^{\prime}$) at high temperature and form gels ($G^{\prime}$ > $G^{\prime\prime}$) with decreasing temperature (Figure~\ref{PicRheology2}A). The gelation temperature ($T_{gel}$) can be identified as a crossover between $G^{\prime}$ and $G^{\prime\prime}$. For the polymer volume fractions considered here, $T_{gel}$ varies between $\approx$60-86~$^\circ$C and increases with $\phi$ (Figure~\ref{PicRheology2}B and Table~S1), as higher endblock concentration facilitates the network formation at higher temperatures.\cite{Mishra2019b} Since $T_{gel}$ is significantly higher than the room temperature (22~$^\circ$C) and the $G^\prime$ displays a plateau at $T<45~^\circ$C for all gels, a well-formed gel microstructure can be expected at room temperature at which most of the experiments have been conducted.

The low-strain gel modulus, as obtained from the plateau modulus in Figure~\ref{PicRheology2}A, increases with $\phi$ (Figure~\ref{PicRheology2}B). A power-law fitting provides $G^{\prime} \sim \phi^{2.39\pm0.38}$. Note that, the $G^{\prime}$ used here corresponds to the oscillation frequency ($\omega$) of 1~rad/s and strain amplitude ($\gamma_0$) of 0.01, the exponent may vary with these parameters. The molecular weight of PI-block in the triblock polymer considered here  ($M_{PI}\approx$79.4~kg/mol) is higher than the PI entanglement molecular weight of $M_{e,PI}\approx$6.4~kg/mol in the melt phase.\cite{rubinstein2003polymer,Chantawansri2013,Mishra2018a} An average number of entanglement per chain in the gel phase can be theoretically estimated by $n_e=\phi^{4/3}(M_{PI}/M_{e,PI})$.\cite{rubinstein2003polymer,Chantawansri2013} For $\phi$=0.0181, the highest polymer volume fraction considered here, we obtain $n_e\approx1.3$ indicating slightly entangled PI-blocks.\cite{Mishra2018a} For other gels, $n_e<1$ suggests no PI entanglements in the gel. Although $\phi^{2.39}$ signifies some level of chain entanglement, as commonly accepted for the chemically crosslinked networks,\cite{rubinstein2003polymer} it is unlikely for all $\phi$ values considered here. In literature, a similar power-law exponent has also been reported for acrylic gels consisting of poly(methylmethacrylate)- poly(n-butyl acrylate)- poly(methylmethacrylate) [PMMA-PnBA-PMMA] in 2-ethyl 1-hexanol, where the midblock entanglement is also unlikely.\cite{Rattan2019} We hypothesize strong  $\phi$ dependence of modulus in our gel is related to the network structure, as at low $\phi$ the loop formation of midblocks is preferred over the bridges. As a result, the aggregates are not well-connected, having some similarities with the colloidal gels for which a power exponent, as high as 4.2 is observed.\cite{Shih1990,Macosko1994,Tadros1996}

We have probed the gel microstructure using small-angle scattering experiments (SAXS) at room temperature. The scattering profiles are typical to self-assembled gels that consist of a combination of structure and form factors related to the spherical aggregates (see Figure~S1A). The structure factor originates from the inter-aggregate scattering and the form factor signifies the size and shape of aggregates. As $\phi$ increases from 0.089 to 0.181, a clear correlation peak related to the inter-aggregate scattering shifts from $q\approx$0.018~\AA$^{-1}$ to 0.02~\AA$^{-1}$ and the intensity ($I(q)$) also increases. Similarly, the second peak related to the form factor shifts from $q\approx$0.031~\AA$^{-1}$ to 0.036~\AA$^{-1}$ (Figure~S1B). A polydispersed core hard-sphere model with Percus-Yevick closure (Figure~S1C) fits the SAXS data reasonably well over the $q$ range of 0.015-0.06~\AA$^{-1}$ (Figure~S1A) validating the micellar microstructure of these gels.\cite{Seitz2007,Zabet2017,Mishra2018a,Mishra2019b} In this model, the PS aggregates represent the spherical core.

Fitting of scattering data captures an increase in the aggregate radius ($r_0$) and a decrease in the hard-sphere thickness ($s$) with increasing $\phi$ (Table~S2). An increase in core radius suggests a higher number of PS-blocks in an aggregate. A decrease in $s$ represents a decrease in distance between the aggregates. The lower hard-sphere radius ($r_0+s$) at higher $\phi$ indicates a relatively compact packing resulting in higher hard-sphere volume fraction ($\psi$). Twice of the hard-sphere thickness ($2s$) provides an estimate of the PI-block length in gel, which we have estimated as 23.8, 20.8, and 17.4~nm for $\phi$=0.089, 0.135, and 0.181, respectively. For a $\theta$-solvent, the end-to-end distance of the PI block can be estimated as, $R_e=b_{PI}\sqrt{N}=21.6$~nm, where, $b_{PI}$ is PI-block Kuhn length, and $N$ is the number of Kuhn segments.\cite{rubinstein2003polymer, Mishra2018a} The ratio of $2s/R_e$ indicates whether the chains are stretched or compressed. We have estimated $2s/R_e$  $\approx$1.1, 0.96, and 0.80 for $\phi$=0.089, 0.135, and 0.181, respectively. These indicate slightly compressed PI-blocks for $\phi$=0.135 and 0.181. In reality, mineral oil is a good solvent for the PI blocks, and the actual $R_e$ is higher than that estimated considering $\theta$-solvent, therefore, the midblocks can be expected to be more compressed. The $2s/R_e>1$ values indicate that the midblocks are slightly stretched at the lower $\phi$.\cite{Mishra2019b} Such stretching of midblocks also promotes the loop formation rather than bridges between the aggregates.

\begin{figure}
	\begin{center}
		\includegraphics[width=3in]{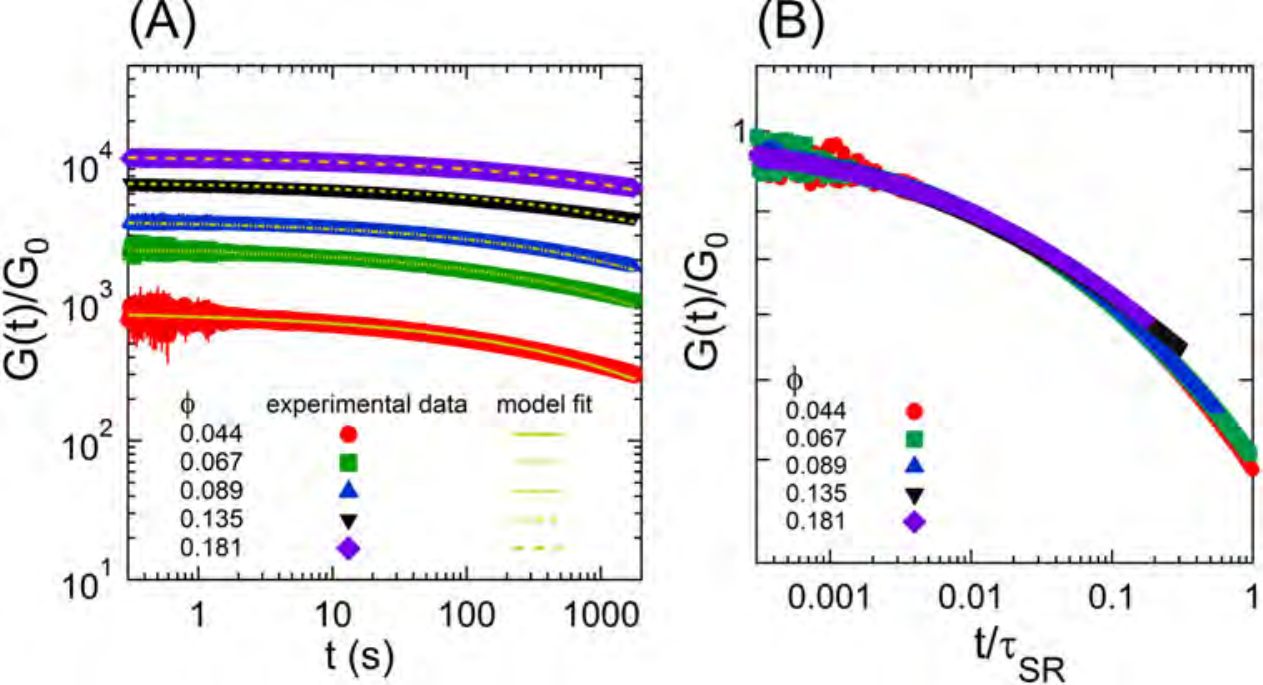}
		%\vspace{-3em}
		\caption{(A)Time dependent shear modulus ($G(t)$) as a function of time ($t$) for all samples obtained from the stress relaxation experiments performed at 22~$^{\circ}$C. A shear strain of 0.01 was used and sample was allowed to relax for 1800~s. (B) Superimposed normalized modulus ($G(t)/G_0$) as a function  of normalized time ($t/\tau_{SR}$) for different  $\phi$. Error bars represent one standard deviation. }
		\label{PicStressrelax}
	\end{center}
\end{figure}

The stress-relaxation in these gels takes place via. endblocks exchange among aggregates.\cite{Seitz2007} The heterogeneous microstructure of these gels results in a range of time scale associated with the endblock exchange, consequentially, these gels exhibit a distribution of relaxation time.\cite{Erk2012a,Zabet2015} Such relaxation behavior is often captured by fitting a stretched exponential model to the stress-relaxation experimental data.\cite{Johnston2006,Seitz2007,Zabet2015} This model is expressed as, $G(t) = G_0exp\left(-\left(t/\tau_{SR} \right)^\beta\right)$, where, $0\le\beta\le1$, $G(t)$ is time-dependent shear modulus, $G_0$ represents the zero-strain shear modulus,\cite{Erk2012a} $\tau_{SR}$ is the characteristic relaxation time, and $\beta$ is the stretched exponent originates from polydispersity in the gel network. A reasonable fitting is obtained for all gels by fixing $\beta$=1/3 (Figure~\ref{PicStressrelax}A).\cite{Erk2012a} $G_0$ and $\tau_{SR}$ have been found to increase with $\phi$ (Table~\ref{TableConcStressRelaxation}). Figure~\ref{PicStressrelax}B displays the normalized stress-relaxation modulus ($G(t)/G_0$) as a function of normalized time ($t/\tau_{SR}$). Interestingly, all relaxation data overlap forming a master curve indicating the same relaxation mechanism in these gels as a result of similar micellar microstructure for all volume fractions.\cite{Erk2012a}

\begin{table}
	\begin{center}
		\caption{Zero-strain shear modulus ($G_0$) and stress relaxation time ($\tau_{SR}$) for samples obtained from fitting the stress relaxation curves to the stretched exponential function.} 
		\label{TableConcStressRelaxation}
		\begin{tabular}{lll}
			\hline
			\multicolumn{1}{p{1.5cm}}{\centering $\phi$}  &  
			\multicolumn{1}{p{2.5cm}}{\centering $G_0(kPa)$} & \multicolumn{1}{p{2.5cm}}{\centering $\tau_{SR}(s)$} \\ 
			
			\hline
			\multicolumn{1}{p{1.5cm}}{\centering 0.044} & \multicolumn{1}{p{2.5cm}}{\centering 0.855$\pm$0.009} & \multicolumn{1}{p{2.5cm}}{\centering 1222$\pm$149} \\ 
			
			\multicolumn{1}{p{1.5cm}}{\centering 0.067} & \multicolumn{1}{p{2.5cm}}{\centering 2.506$\pm$0.017} & \multicolumn{1}{p{2.5cm}}{\centering 1835$\pm$146} \\ 
			
			\multicolumn{1}{p{1.5cm}}{\centering 0.089} & \multicolumn{1}{p{2.5cm}}{\centering 3.863$\pm$0.009} & \multicolumn{1}{p{2.5cm}}{\centering 3158$\pm$101} \\ 
			
			\multicolumn{1}{p{1.5cm}}{\centering 0.135 } & \multicolumn{1}{p{2.5cm}}{\centering 7.356$\pm$0.011} & \multicolumn{1}{p{2.5cm}}{\centering 5765$\pm$129} \\
			
			\multicolumn{1}{p{1.5cm}}{\centering  0.181 } & \multicolumn{1}{p{2.5cm}}{\centering 11.183$\pm$0.008} & \multicolumn{1}{p{2.5cm}}{\centering 10163$\pm$131} \\ 
			
			\hline
		\end{tabular}
	\end{center}
\end{table}

Below $T_{gel}$, the time-temperature-superposition of the rheological data can be achieved to obtain a master curve for  $G^{\prime}$ and $G^{\prime\prime}$ as a function of shifted oscillation frequency ($a_T \omega$), where $a_T$ is a shifting factor (Figure~S2).\cite{Peters2016,Mishra2019b} The $a_T$ values are fitted with Arrhenius equation, $a_T=\exp\left((E_a/R)(1/T-1/T_{ref})\right)$, where $R$ is gas constant and $T_{ref}$=22~$^\circ$C is the reference temperature. $E_a$ is a fitting parameter representing the activation energy mostly related to the endblock pull-out or endblock exchange.\cite{Inomata2003}  Interestingly, the $E_a$ values do not change significantly with the polymer volume fraction and are in the range of $ \approx$185-215~kJ/mol (Table~S3). This further supports our argument made from the stress relaxation experiments that the relaxation process in all gels is similar and endblock pullout from aggregates dominates the relaxation process. Also, such a relaxation process does not change with temperature, as long as the samples are in the gel state, i.e., below the gelation temperature.

\begin{figure*}
	\begin{center}
		\includegraphics[width=5.5in]{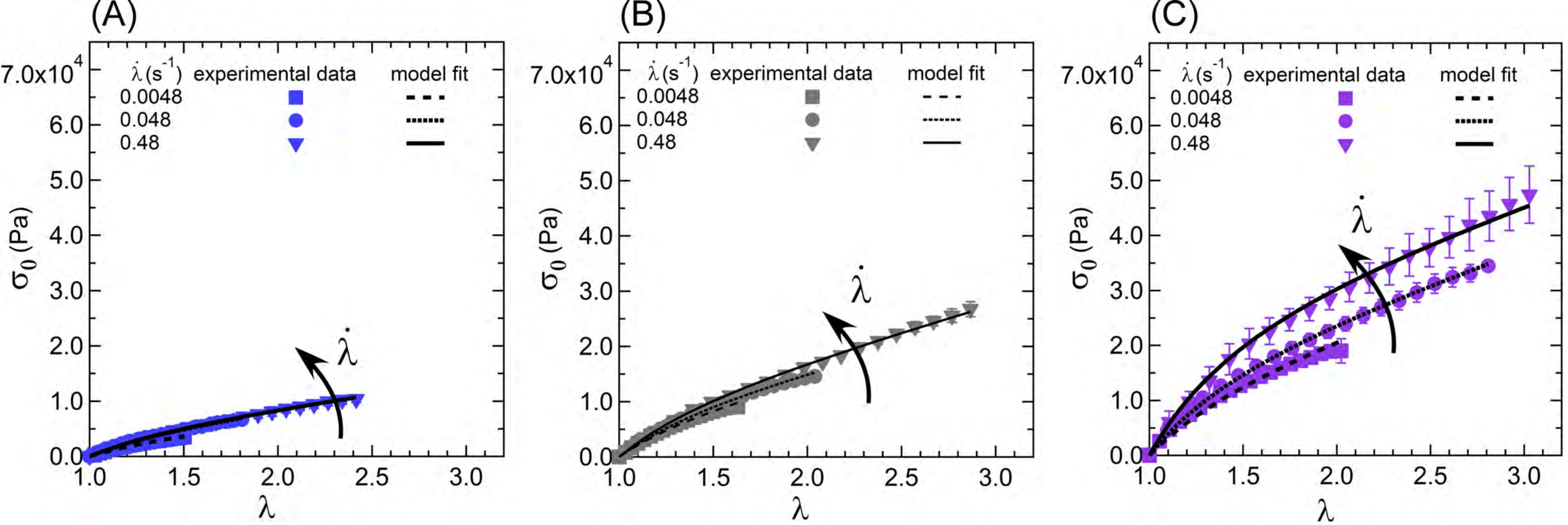}
		%\vspace{-3em}
		\caption{Tensile test results for $\phi$=  0.089 (A),  0.135 (B), and  0.181 (C) performed at $\dot{\lambda}\approx$ 0.0048, 0.048 and 0.48~s$^{-1}$, respectively. Error bars represent one standard deviation. The line represent the model fit for each case. }
		\label{PicConcTensileGraphs}
	\end{center}
\end{figure*}

We have not observed any significant variation in $G^{\prime}$ and $G^{\prime\prime}$ while varying the oscillation amplitude, 10$^{-4}<\gamma_0<$1 (Figure~S3). To probe the large deformation behavior of these gels, we have conducted tensile tests. Figure~\ref{PicConcTensileGraphs}A-C displays nominal stress ($\sigma_0$) as a function of stretch ratio ($\lambda$) obtained from the tensile tests performed on gels with $\phi$=0.089, 0.135, and 0.181 using stretch rate $\dot{\lambda}$ =0.0048, 0.048, and 0.48~s$^{-1}$, respectively. Due to the low sample stiffness, tensile tests could not be performed on the gels with $\phi$=0.044 and 0.067. For $\phi$=0.089 and 0.135, all three $\dot{\lambda}$-curves overlap indicating that the effect of $\dot{\lambda}$ is not significant, while for $\phi$=0.181, all $\dot{\lambda}$-curves are distinct (Figure~\ref{PicConcTensileGraphs}).

For $\phi$=0.089 and 0.135, the low-strain modulus can be estimated by fitting neo-Hookean model with the experimental data. For the uniaxial load, $\sigma_{0}=  G_c\left(\lambda-\lambda^{-2}\right)$, where $G_c$ is shear modulus contributed by the crosslinks.\cite{rubinstein2003polymer} We obtain $G_c\approx4.3\pm$1 and $8.6\pm$1~kPa for $\phi$=0.135 and 0.089, respectively. These results are similar to that obtained from the temperature sweep experiments (Figure~\ref{PicRheology2}A and Table~S1). For  $\phi$=0.181, an increase of modulus ($d\sigma_0/d\lambda$) with increasing $\dot{\lambda}$ can be observed (Figure~\ref{PicConcTensileGraphs}), which can be attributed to the slight entanglement of the PI-blocks.\cite{Mishra2018a,Mishra2019b} Slip and Tube model\cite{Rubinstein2002} expressed as, $\sigma_0=\left[G_c+G_e(0.74 \lambda + 0.16 \lambda^{-1/2} - 0.35)^{-1}\right](\lambda-\lambda^{-2})$, captures the data reasonably well. The fitting provides the contribution of  entanglements in modulus ($G_e$) as 0.079, 2.7 and 8.7~kPa for $\dot{\lambda}$= 0.0048, 0.048 and 0.48~s$^{-1}$, respectively, while fixing the $G_c=G^{\prime}_{22^\circ C}$=11.1~kPa.\cite{Mishra2018a} This suggests that at higher $\dot{\lambda}$, the entanglements behave like temporary crosslinks and increase the modulus of gels. An increase in sample stretchability, i.e., higher failure strain ($\lambda_f$), at higher $\phi$ is observed (Figures~S4 and~S5) and that can be attributed to the slightly compressed midblocks at higher $\phi$, resulting in a higher stretch before failure. At higher $\dot{\lambda}$, the PI-chains stretch faster, not allowing sufficient time for stress transfer to the PS-blocks for pullout from aggregates to take place and this likely leads to a higher $\lambda_f$. 

\begin{figure*}
	%	\begin{center}
	\includegraphics[width=5.5in]{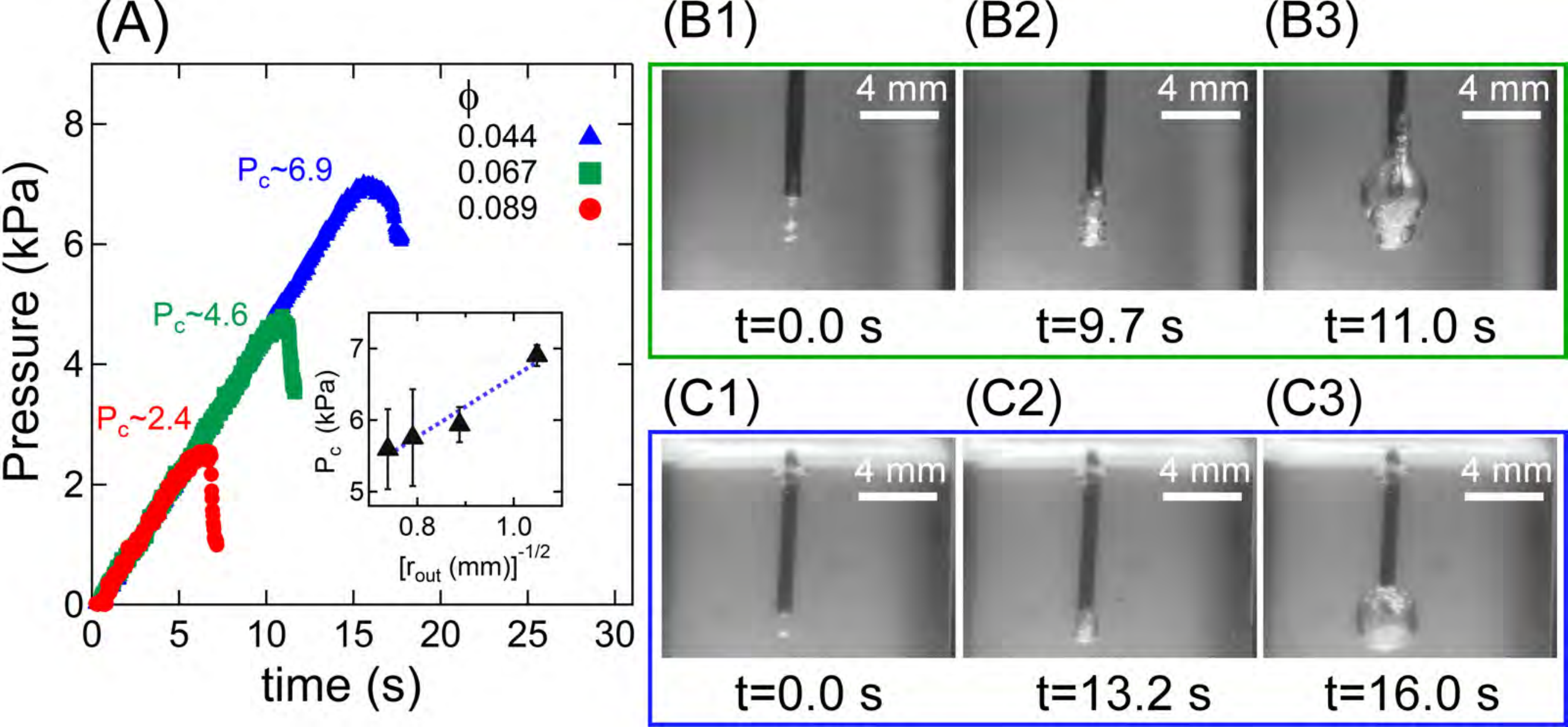}
	%\vspace{-3em}
	\caption{(A) Pressure as a function of time from the cavitation rheology (CR) experiments for the polymer volume fractions, $\phi$=0.044, 0.067, and 0.089, respectively. Inset represents the critical pressure ($P_c$) as a function of needle outer radius ($r_{out}$). Photomicrographs of cylindrical cavity expansion for  $\phi$=0.067 (B1-B3), and  $\phi$=0.089 (C1-C3).}
	\label{PicConcCavitationGraph}
	%	\end{center}
\end{figure*}

To understand the effect of $\phi$ on the fracture behavior, we have determined the critical energy release rate ($\Gamma_0$) for these gels by selectively conducting cavitation rheology (CR), and fracture experiments initiated from a predefined crack.  Because of low-modulus, gels with $\phi$=0.044 and 0.067 cannot be tested using fracture experiments, therefore, only CR has been used. For $\phi$=0.089, 0.135 and 0.181, fracture experiments have been conducted. CR has also been performed on $\phi$=0.089 to compare $\Gamma_0$ obtained from both of these experiments. 
For $\phi$=0.135 and 0.181, weak adhesion between the gel and needle-wall was observed resulting in air escape through the interface. Therefore, CR experiments for these samples are not reliable. However, a combination of CR and fracture experiments enable us to investigate the failure of these gels over a wider concentration range.

Inserting a needle in a gel can pre-stress the gel in front of the needle-tip. However, it has been shown that retracting the needle beyond a critical distance ($d_r$) can release that stress and the subsequent CR experiments provide reasonable results.\cite{Barney2019} Now, $d_r$ can be estimated as $d_r\approx\sqrt{r_{out}\Gamma_0/G^{\prime}} \approx$1.5~mm, by considering $\Gamma_0\approx$50~J/m$^2$, $G^{\prime}\approx$ 10$^4$~Pa, and $r_{out}\approx$454~$\mu$m. In our experiments, we have used  $d_r$$\approx$3.75~mm, enough to reduce the impact of needle-induced stress on the critical pressure. For the needle with $r_{out} \approx$454~$\mu$m, the needle retraction creates a hollow cylinder or cavity with the aspect ratio, $L/r_{out}\approx8.25$.  Upon pressurization beyond a critical pressure ($P_c$), this cavity expands in the radial direction because of elastic instability or fracture of the gel.

Figure~\ref{PicConcCavitationGraph}A displays the applied pressure as a function of time for $\phi$=0.044, 0.067, and 0.089, respectively. The cavity expansion during pressurization for $\phi$=0.067 and 0.089 are shown in  Figure~\ref{PicConcCavitationGraph}B1-B3 and C1-C3, respectively. Interestingly, $P_c$ is not very sharp in the present gel in comparison to that observed for polyacrylamide gels,\cite{Kundu2009} vitreous biomaterials,\cite{Zimberlin2007} and acrylic gels.\cite{Hashemnejad2015} Rather, the pressure drop appears to be smoother, which is similar to that reported for the high concentration cured PDMS (Sylgard 184) and has been attributed to the gel fracture.\cite{Barney2019} 

For crack growth in an incompressible linear elastic material, $\Gamma_{0,LE}\approx (3\pi/4)(P_c^2r_{out}/ E)$ considering plain strain approximation, where $P_c$ corresponds to the critical pressure for gel fracture (see Supporting Information).\cite{anderson2017fracture} Using the approach, we obtain $\Gamma_{0,LE}\approx$~3.17, 4.47, and 5.66~J/m$^2$ for $\phi$=0.044, 0.067, and 0.089, respectively (see Table~S4). While for the inflation of well-defined cylindrical shaped crack, an approach provided by William and Schapery (WS)\cite{Williams1965} estimates $\Gamma_{0,WS} \approx (27/4)(P_c^2r_{out}/ E)$ (Table~S4). Our calculation indicates $\Gamma_{0,WS}\approx$~9.10,12.80, and 16.24~J/m$^2$ for $\phi$=0.044, 0.067, and 0.089, respectively (Table~S4). Both of the above approaches captures $P_c \sim r_{out}^{-1/2}$ relationship. As shown in Figure~\ref{PicConcCavitationGraph}A inset, this relationship has been validated by using needles of different radius as we obtain a reasonably good linear fit between $P_c$ and $r_{out}^{-1/2}$.

\begin{figure}
	%	\begin{center}
	\includegraphics[width=2.5in]{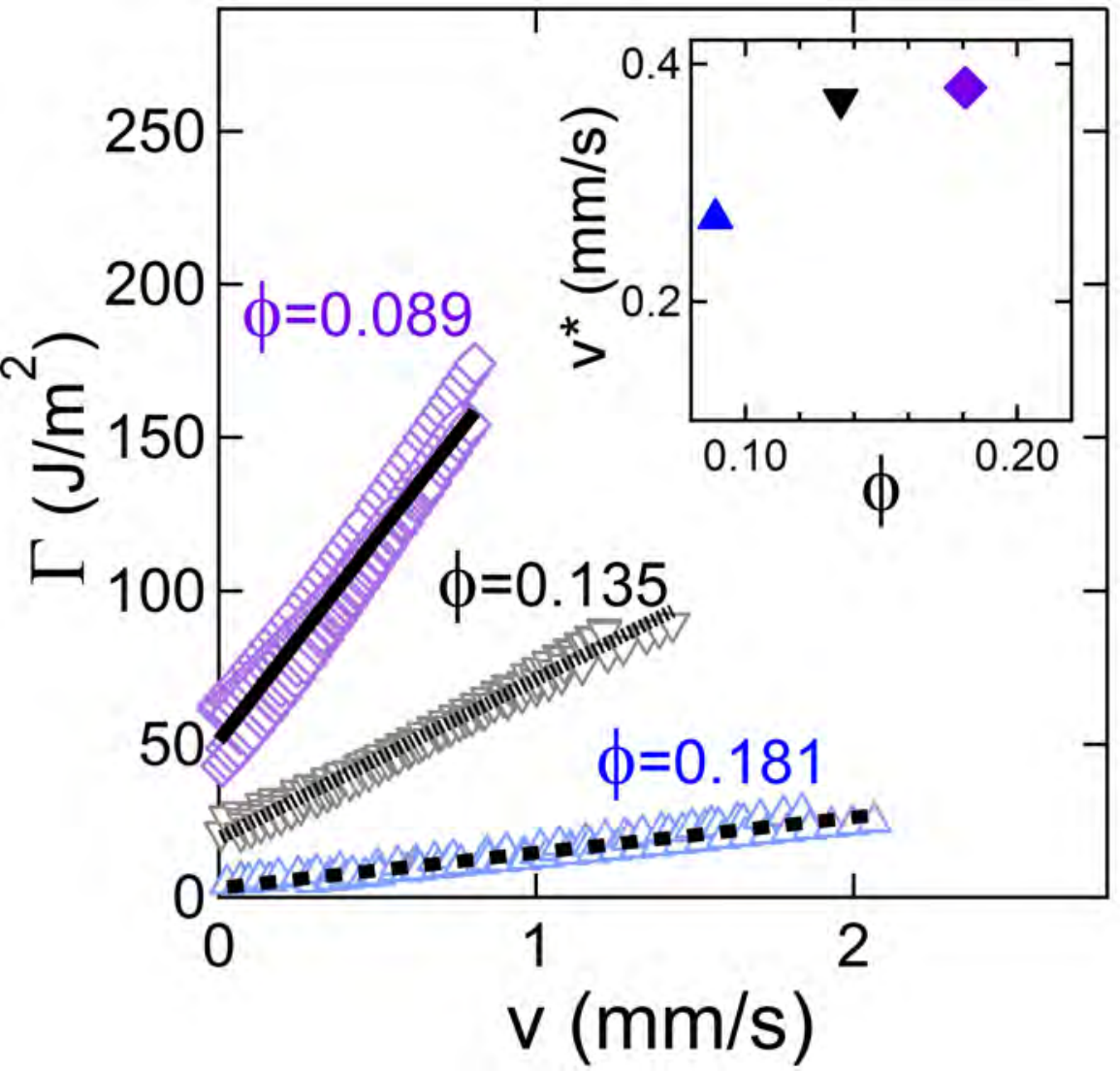}
	%\vspace{-3em}
	\caption{Energy release rate ($\Gamma$) as a function of crack tip velocity ($v$) obtained from the fracture experiments with a predefined crack for $\phi$=0.089, 0.135, and 0.181, respectively. The inset shows the critical crack velocity ($v^*$) as a function $\phi$.}
	\label{PicConcFractureGraph}
	%	\end{center}
\end{figure}

Both of the approaches estimate the release of strain energy stored in the material upon crack opening, however, it differs due to the initial shape of the cavity. The prefactor 27/4 in the linear elasticity approach is dictated by the stress concentration (localization) in the crack geometry, while the WS approach does not consider such stress concentration.  As a result, the $\Gamma_{0,LE}$ is 2.8 times lower than $\Gamma_{0,WS}$, similar to what observed in the case of penny shape crack.\cite{Lin2004} In our case, the needle retraction creates a distorted cylinder with defects at the wall (see Figures~\ref{PicConcCavitationGraph}B1 and C1) leading to the stress-localization at those defects, therefore, the linear elastic approach is more relevant for our case.

The results of fracture experiments with a predefined crack are displayed in Figure~\ref{PicConcFractureGraph} for $\phi$=0.089, 0.135, and 0.181, respectively. In these experiments, a rectangular sample with a sharp crack is clamped between two bars. The top bar has been moved upward continuously to extend the crack length. The energy release rate ($\Gamma$) is calculated from the energy stored in the sample at a particular crack length as a result of applied stretch.\cite{Baumberger2006,Seitz2009,Mishra2018a} The crack-tip velocity ($v$) is estimated from the time-lapsed images of crack propagation.\cite{Mishra2018a} For all samples $\Gamma$ scales linearly with the crack-tip velocity ($v$),  indicating viscous-dissipation phenomena related to the PI-chain motion in mineral oil.\cite{Baumberger2006,Mishra2018a} The experimental data is fitted with the expression, $\Gamma = \Gamma_0(1+v/v^\ast)^n$, where $v^*$ is the characteristic crack tip velocity, and exponent $n$ is a fitting parameter fixed to 1 due to the linear relationship between $\Gamma$ and $v$.\cite{Mishra2018a, Mishra2019b} We obtain $v^\ast$ varying between 0.27-0.38~mm/s and decreases slightly with increasing $\phi$ (Figure~\ref{PicConcFractureGraph} inset and Table~\ref{TableConcFractureFittingResults}). The $\Gamma_0$ values for $\phi$=0.089, 0.135, and 0.181 have been estimated as 3.10, 19.27, and 51.25~J/m$^2$, respectively. $\Gamma_0$ for $\phi$=0.089 obtained from the CR and fracture experiments are of the same order of magnitude, even considering different possible models to estimate $\Gamma_0$ from the $P_c$ value. There are other factors that can contribute to the different $\Gamma_0$ values between two techniques such as uncertainties in detecting the exact crack-tip position, and system compliance for the fracture and CR experiments.

\begin{table}
	\begin{center}
		\caption{Critical energy release rate ($\Gamma_0$) and characteristic crack-tip velocity ($v^\ast$) as fitted parameters obtained from fitting the fracture experiment data using  $\Gamma = \Gamma_0(1+v/v^\ast)^{n=1}$. } 
		\label{TableConcFractureFittingResults}
		\begin{tabular}{lll}
			\hline
			\multicolumn{1}{p{1.5cm}}{\centering $\phi$}  &  
			\multicolumn{1}{p{2.5cm}}{\centering $\Gamma_0$ (J/m$^2$)} &
			\multicolumn{1}{p{2cm}}{\centering $v^{\ast}$ (mm/s)} \\  
			
			\hline
			
			\multicolumn{1}{p{1.5cm}}{\centering 0.089} & \multicolumn{1}{p{2.5cm}}{\centering 3.10 $\pm$ 0.40} & 
			\multicolumn{1}{p{2cm}}{\centering 0.27} \\

			\multicolumn{1}{p{1.5cm}}{\centering 0.135} & \multicolumn{1}{p{2.5cm}}{\centering 19.27 $\pm$ 0.43} & 
			\multicolumn{1}{p{2cm}}{\centering 0.37} \\
			
			\multicolumn{1}{p{1.5cm}}{\centering 0.181} & \multicolumn{1}{p{2.5cm}}{\centering 51.25 $\pm$ 1.01} & 
			\multicolumn{1}{p{2cm}}{\centering 0.38} \\
			
			\hline
		\end{tabular}
	\end{center}
\end{table}

\begin{figure}[b]
	\begin{center}
		\includegraphics[width=2.5in]{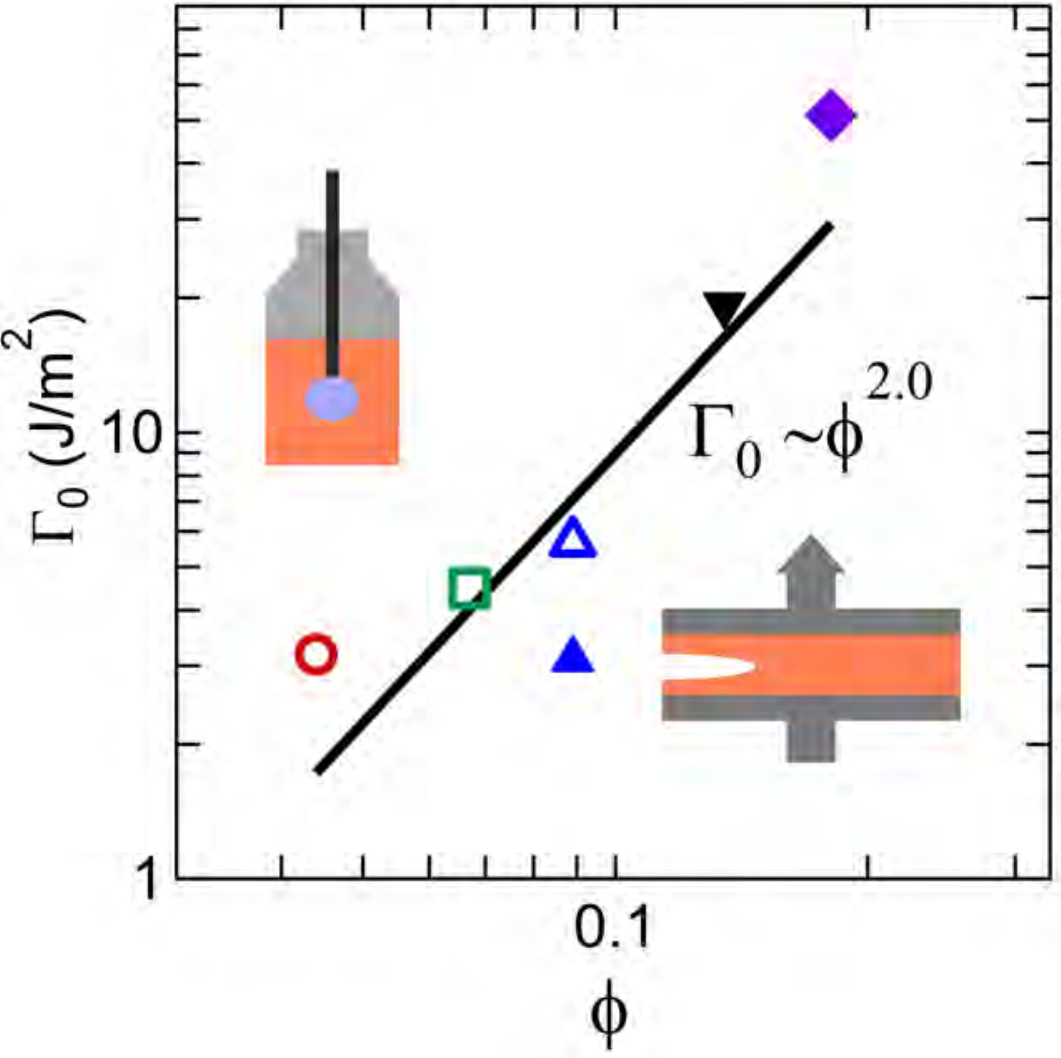}
		%\vspace{-3em}
		\caption{Critical energy release rate ($\Gamma_0$) obtained from the cavitation rheology (open symbols) and fracture experiments (closed symbols) as a function of polymer volume fraction($\phi$). }
		\label{PicConcFractureFit}
	\end{center}
\end{figure}

Figure~\ref{PicConcFractureFit} summarizes the $\Gamma_0$ values obtained from the CR and fracture experiments as a function of $\phi$.  A power-law fit ( $\Gamma_0 \sim \phi^m$) estimates the exponent $m$ as 2.0. For acrylic gels, a similar value of $m$=2.1 has been determined from fracture experiments.\cite{Rattan2019} Similar to the present case, the midblock molecular weight in that study was higher than the entanglement molecular weight of the polymer and no midblock entanglement is expected in the probed range of $\phi$.\cite{Rattan2019} In contrast, using CR experiments on gelatin gels, it has been shown that  $m$=2.4 and 0.6 for the entangled and non-entangled systems, respectively.\cite{Frieberg2018} The $\Gamma_0-\phi$ relationship for physically assembled gels is not completely understood and in our case it likely depends on the  ratio of loop and bridge fractions. However, further investigation is needed.

\section{Conclusions}

Our study develops a comprehensive understanding of small and large deformation mechanical behavior of physically assembled PS-PI-PS gels over a wide $\phi$-range probed by using multiple experimental techniques. A similar relaxation mechanism displayed by a superposition of stress relaxation results can be attributed to the similar micellar microstructure. Due to the entanglement of PI-block at higher volume fraction, the gel displays a strain-rate dependent modulus at higher $\phi$.  The critical energy release rate determined by cavitation rheology and fracture experiments displays $\Gamma_0 \sim \phi^{2.0}$.  $G^\prime \sim \phi^{2.39}$ relationship observed for these gels can be attributed to the change in loop-to-bridge fraction ratio with increasing $\phi$, however, further investigation is needed for a quantitative analysis.\cite{Sato1996}

\section{acknowledgement}
	
	This work was supported by the National Science Foundation [DMR-1352572]. We are grateful to Dr. Xiaodan Gu at the University of Southern Mississippi for conducting SAXS experiments.
	
\pagebreak

\section{Supplementary Information}

	The supporting information is available. The following files are available free of charge.
 Filename: Supporting Information.pdf. \\
 Tables and Figures: additional rheology data (time-temperature superposition and amplitude sweep); Scattering data with model fit; Additional tensile test data;  Estimation of energy release rate: WS approach and linear elastic mechanics approach.

\bibliography{Mishra_el_al_biblography}

\end{document}

\begin{table}[H]
	\begin{center}
		\caption{Gelation temperature ($T_{gel}$), and storage modulus at room temperature ($G^{\prime}_{22^\circ C}$) for all gel samples. }
		\label{TableConcGelRheology}
		\begin{tabular}{lll}
			\hline
			\multicolumn{1}{p{1.7cm}}{\centering $\phi$}  &
			\multicolumn{1}{p{1.7cm}}{\centering $T_{gel}$}  &   
			\multicolumn{1}{p{2.5cm}}{\centering $G^{\prime}_{22^\circ C}$ (kPa)}   \\

			\hline
			\multicolumn{1}{p{1.7cm}}{\centering 0.044}  &
			\multicolumn{1}{p{1.7cm}}{\centering 59.8}  &   
			\multicolumn{1}{p{2.5cm}}{\centering 0.671}  \\

			\multicolumn{1}{p{1.7cm}}{\centering 0.067}  &
			\multicolumn{1}{p{1.7cm}}{\centering 60.7}  &   
			\multicolumn{1}{p{2.5cm}}{\centering 1.929}   \\
			
			\multicolumn{1}{p{1.7cm}}{\centering 0.089}  &
			\multicolumn{1}{p{1.7cm}}{\centering 63.4}  &   
			\multicolumn{1}{p{2.5cm}}{\centering 2.923 } \\

			\multicolumn{1}{p{1.7cm}}{\centering 0.135}  &
			\multicolumn{1}{p{1.7cm}}{\centering 72.3}  &   
			\multicolumn{1}{p{2.5cm}}{\centering 4.913}  \\

			\multicolumn{1}{p{1.7cm}}{\centering 0.181}  &
			\multicolumn{1}{p{1.7cm}}{\centering 86.0}  &   
			\multicolumn{1}{p{2.5cm}}{\centering 12.692}  \\

			\hline
		\end{tabular}
	\end{center}
\end{table}

\begin{figure}[H]
	\begin{center}
		\includegraphics[width=6.5in]{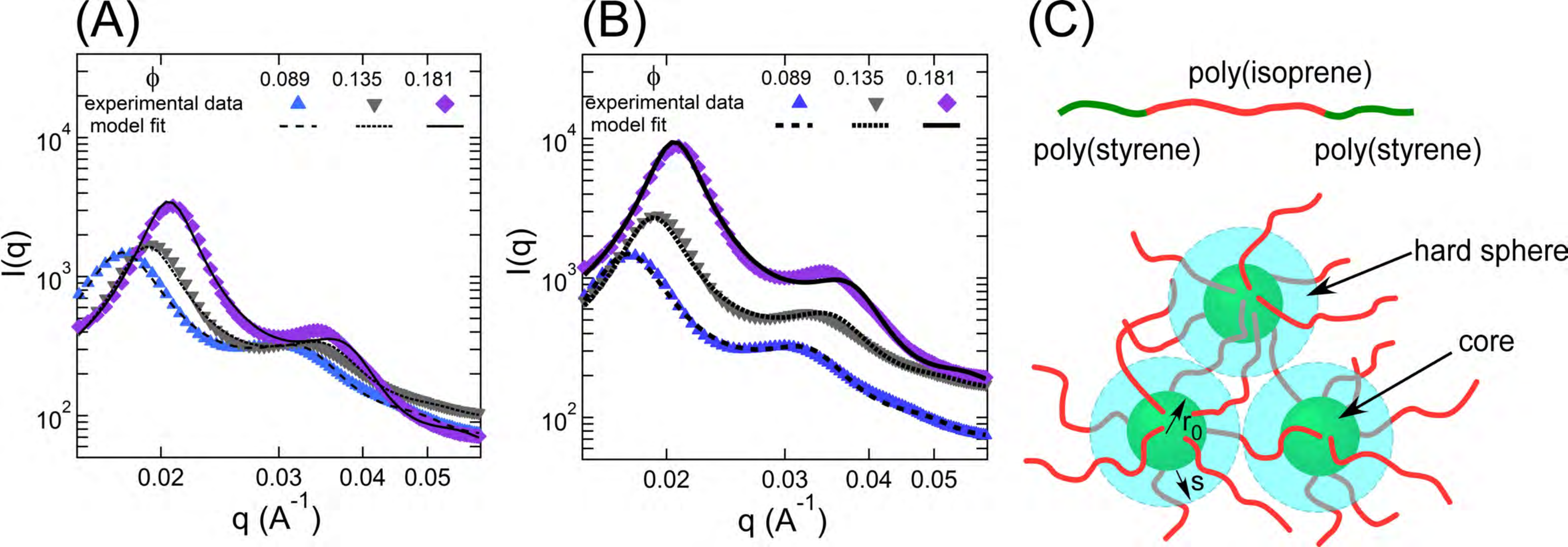}
		%\vspace{-3em}
		\caption{ (A) Circular averaged intensity ($I(q)$) as a function of scattering angle ($q$) for $\phi$=0.089, 0.135, and 0.181, respectively. The lines show fit with the polydispersed core hard-sphere model with a Percus-Yevick approximation. (B) For illustration, the $I(q)$ and corresponding fit line for $\phi$=0.135 and 0.181 are shifted by 3000 and 6000 units along the y-axis, respectively. (C) Schematic representing the polydispersed core hard-sphere model, where the core represents the PS aggregates with a Gaussian distribution ($\sigma$) in radius ($r_{0}$). The hard-spheres are fictitious in nature with thickness of $s$. The hard-sphere volume fraction is $\psi$. }
		\label{PicSupConcSAXS}
	\end{center}
\end{figure}

 \begin{table}[H]
 	\caption{Parameters obtained from fitting small angle scattering (SAXS) data to polydispersed core hard-sphere model with Percus-Yevick closure.}
 	\label{TableSupConcSAXSResults}
 	\begin{tabular}{llllll}
 		\hline
 		\multicolumn{1}{p{1.7cm}}{\centering $\phi$}  &   		\multicolumn{1}{p{2.5cm}}{\centering Core radius,\\ $r_{0}$(nm)} &  
 		\multicolumn{1}{p{2.5cm}}{\centering Hard-sphere \\thickness,$s$(nm)} & 
 		\multicolumn{1}{p{2cm}}{\centering Polydispersity\\($\sigma/r_{0}$)} & 
 		\multicolumn{1}{p{2.5cm}}{\centering Volume\\ fraction($\psi$)} & 
 		\multicolumn{1}{p{2cm}}{\centering Aggregation \\number}\\

 		\hline
 		\multicolumn{1}{p{1.7cm}}{\centering 0.089} & \multicolumn{1}{p{2.5cm}}{\centering 7.1} & \multicolumn{1}{p{2.5cm}}{\centering 11.9} & \multicolumn{1}{p{2cm}}{\centering 0.29}  &  \multicolumn{1}{p{2.5cm}}{\centering 0.42}  & 
 		\multicolumn{1}{p{2cm}}{\centering 62}\\

 		\multicolumn{1}{p{1.7cm}}{\centering 0.135} & \multicolumn{1}{p{2.5cm}}{\centering 7.2} & \multicolumn{1}{p{2.5cm}}{\centering 10.4} & \multicolumn{1}{p{2cm}}{\centering 0.19} &   \multicolumn{1}{p{2.5cm}}{\centering 0.44} & 
 		\multicolumn{1}{p{2cm}}{\centering 71}\\
 		
 		\multicolumn{1}{p{1.7cm}}{\centering 0.181} & \multicolumn{1}{p{2.5cm}}{\centering 8.0} & \multicolumn{1}{p{2.5cm}}{\centering 8.7} & \multicolumn{1}{p{2cm}}{\centering 0.24} &   \multicolumn{1}{p{2.5cm}}{\centering 0.47} & 
 		\multicolumn{1}{p{2cm}}{\centering 76}\\
 		\hline
 	\end{tabular}
 \end{table}

\begin{center}
         	
       \includegraphics[trim={0 2in 0 0}, clip,width=4.0in]{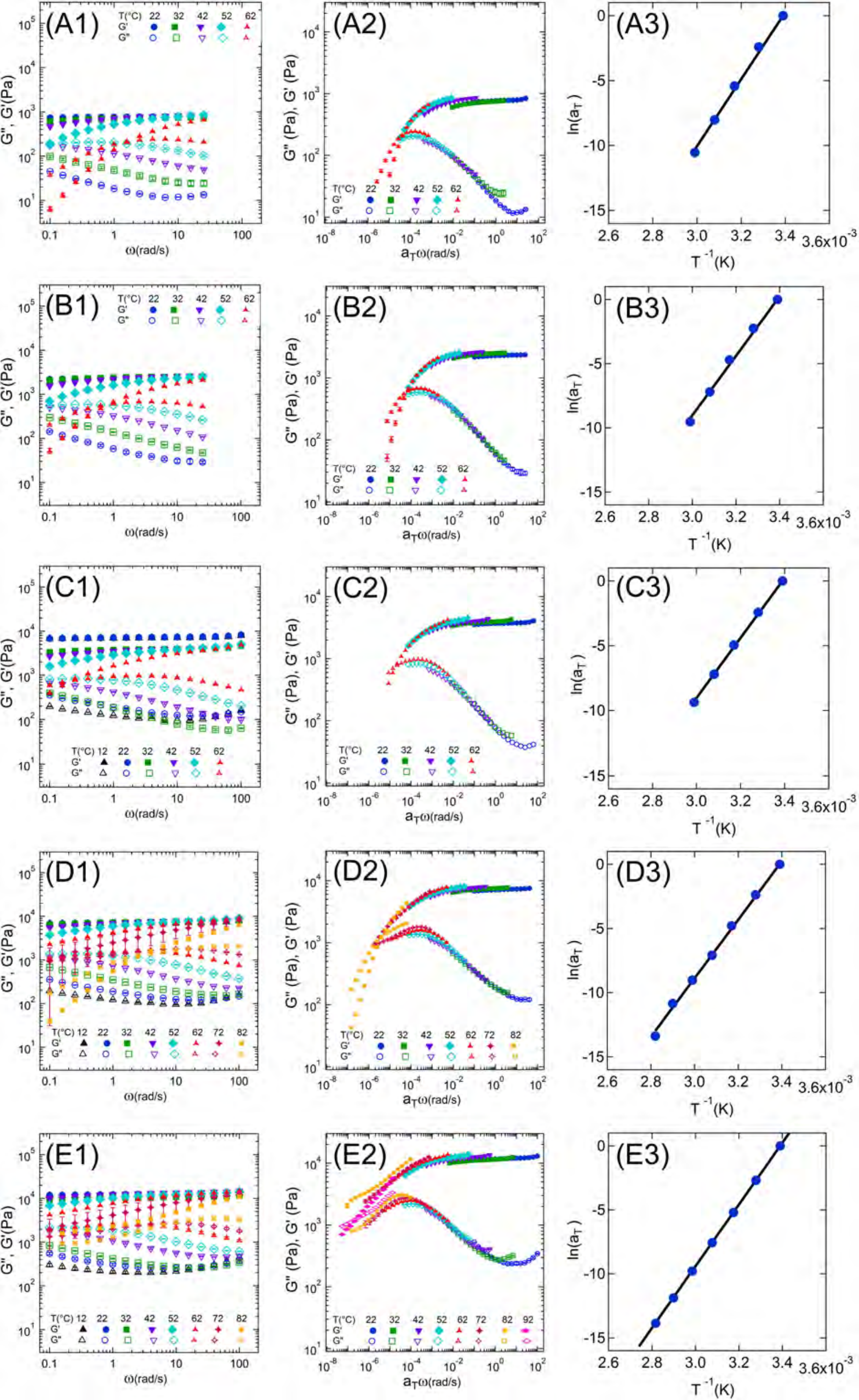}
       \includegraphics[trim={0 0 0 8.2in}, clip,width=4.0in]{PicTTS5_All.pdf}
        
    \end{center}
\begin{figure}[H]

	\caption{(A1) Evolution of $G^{\prime}$ and $G^{\prime\prime}$ as a function of frequency ($\omega$) using strain amplitude $\gamma_0$=0.01 for $\phi$=  0.044(A1), 0.067(B1),  0.089(C1), 0.089(D1), and  0.181(E1), respectively, at different temperatures. The time-temperature superposition (TTS) curves for $\phi$=  0.044(A2),  0.067(B2),  0.089(C2),  0.089(D2), and  0.181(E2), respectively, obtained by shifting the curves horizontally using a shifting factor $a_T$. Shift factor $a_T$ fitted with Arrhenius equation for  $\phi$=  0.044(A3),  0.067(B3),  0.089(C3),  0.089(D3), and  0.181(E3), respectively, using the reference temperature $T_{ref}$=22~$^\circ$C. Error bars represent one standard deviation.} 
	\label{PicSupConcFreqSweep}
\end{figure}

 \begin{table}[H]
 	\begin{center}
 		\caption{Activation energy ($E_a$) for all samples obtained by fitting Arrhenius equation  with the TTS shift factors ($a_T$). Here, the reference temperature is $T_{ref}$=22~$^\circ$C. } 
 		\label{TableSupTTSFit}
 		\begin{tabular}{ll}
 			\hline
 			\multicolumn{1}{p{1.7cm}}{\centering $\phi$}  &
 			\multicolumn{1}{p{2.3cm}}{\centering  $E_a$ (kJ/mol)}  \\  
 			%	\multicolumn{1}{p{2.0cm}}{\centering $\omega_{co}$ (rad/s)}   &
 			%	\multicolumn{1}{p{1.7cm}}{\centering  $\tau_{FS}$ (s)}  \\ 

 			\hline
 			\multicolumn{1}{p{1.7cm}}{\centering 0.044}  &
 			\multicolumn{1}{p{2.3cm}}{\centering 210.9}  \\  
 			%	\multicolumn{1}{p{2.0cm}}{\centering 4.40$\times$10$^{-5}$}   &
 			%	\multicolumn{1}{p{1.7cm}}{\centering 2.27$\times$10$^{4}$}  \\ 

 			\multicolumn{1}{p{1.7cm}}{\centering 0.067}  &
 			\multicolumn{1}{p{2.3cm}}{\centering 195.8}  \\ 
 			%	\multicolumn{1}{p{2.0cm}}{\centering 5.14$\times$10$^{-5}$}    &
 			%	\multicolumn{1}{p{1.7cm}}{\centering 1.45$\times$10$^{4}$}  \\ 
 			
 			\multicolumn{1}{p{1.7cm}}{\centering 0.089}  &
 			\multicolumn{1}{p{2.3cm}}{\centering 191.6}  \\  
 			%	\multicolumn{1}{p{2.0cm}}{\centering 1.69$\times$10$^{-5}$ }  &
 			%	\multicolumn{1}{p{1.7cm}}{\centering 5.88$\times$10$^{4}$}  \\ 

 			\multicolumn{1}{p{1.7cm}}{\centering 0.135}  &
 			\multicolumn{1}{p{2.3cm}}{\centering 186.8}  \\  
 			%	\multicolumn{1}{p{2.0cm}}{\centering 0.39$\times$10$^{-5}$}   &
 			%	\multicolumn{1}{p{1.7cm}}{\centering 25.5$\times$10$^{4}$}  \\ 

 			\multicolumn{1}{p{1.7cm}}{\centering 0.181}  &
 			\multicolumn{1}{p{2.3cm}}{\centering 215.6}  \\  
 			%	\multicolumn{1}{p{2.0cm}}{\centering 0.02$\times$10$^{-5}$}   &
 			%	\multicolumn{1}{p{1.7cm}}{\centering 476$\times$10$^{4}$}  \\ 
 			
 			\hline
 		\end{tabular}
 	\end{center}
 \end{table}

\begin{figure}[H]
	\begin{center}
		\includegraphics[ width=3in]{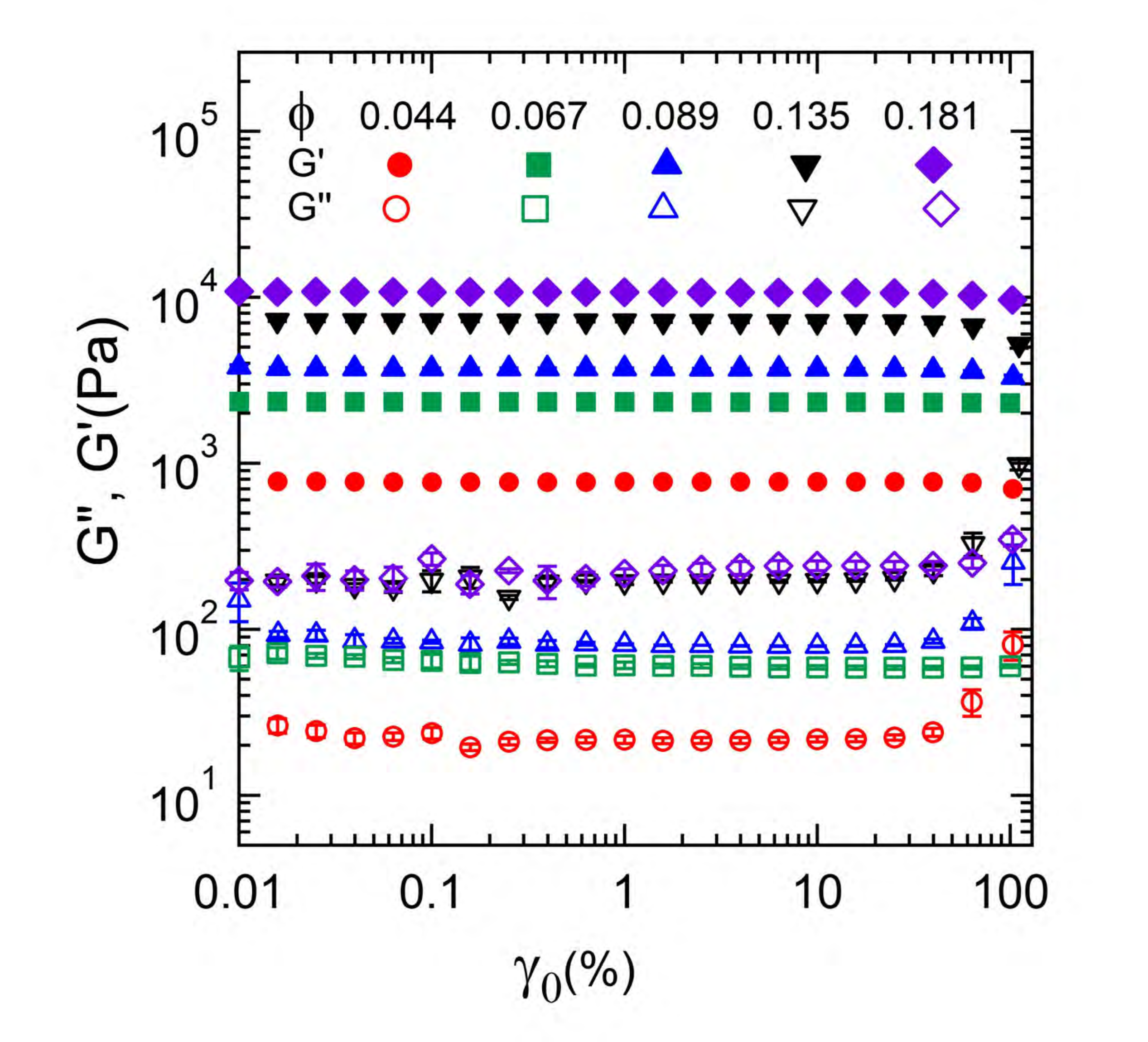}
		%\vspace{-3em}
		\caption{$G^{\prime}$ and $G^{\prime\prime}$ plotted for $\phi$=0.044, 0.067, 0.089, 0.135, and 0.181, respectively, as a function of oscillation amplitude, $\gamma_0$=0.01-100\% using oscillation frequency, $\omega$=1~rad/s at 22~$^{\circ}$C. Error bars represent one standard deviation. }
		\label{PicSupLAOSFreqSweep}
	\end{center}
\end{figure}

%\begin{figure}[H]
%	\begin{center}
%		\includegraphics[width=3in]{PicGelationTemp.pdf}
%		%\vspace{-3em}
%		\caption{Gelation temperature ($T_{gel}$) as a function of polymer volume fraction ($\phi$). }
%		\label{PicConcGelationTPhi}
%	\end{center}
%\end{figure}

%\begin{figure}[H]
%	\begin{center}
%		\includegraphics[width=4in]{Pic20pTTS.pdf}
%		%\vspace{-3em}
%		\caption{$G^{\prime}$ and $G^{\prime\prime}$ plotted as a function of shifted frequency ($a_T\omega$) by using the time temperature superposition (TTS) for $\phi$=0.181. Experiments were performed at $\omega$= 1~rad/s and $\gamma_0$ =0.01 and curves are shifted using the reference temperature, $T_{ref}$=$22~^{\circ}$C. Error bars represent one standard deviation.} 
%		\label{PicConcFreqSweep20p}
%	\end{center}
%\end{figure}
\begin{figure}
	\begin{center}
		\includegraphics[width=3in]{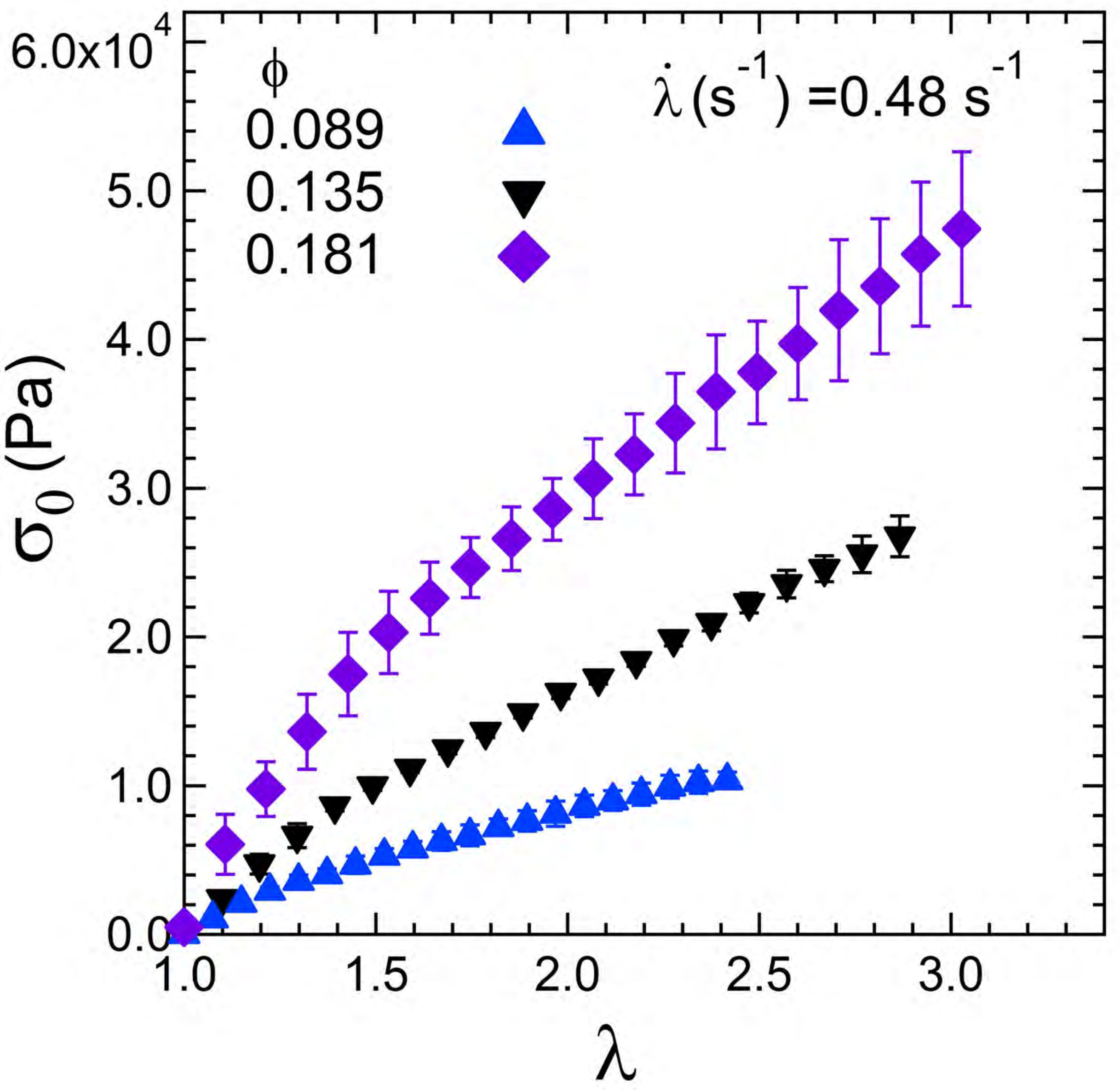}
		%\vspace{-3em}
		\caption{Nominal stress ($\sigma_0$) as a function of stretch ratio ($\lambda$) obtained from the tensile tests using stretch rate,  $\dot{\lambda}\approx$0.48~s$^{-1}$ for  $\phi$=0.089, 0.135, and 0.181, respectively. Error bars represent one standard deviation.}
		\label{PicSupTensileHighStretchRate}
	\end{center}
\end{figure}

\begin{figure}[H]
\begin{center}
		\includegraphics[width=3.0in]{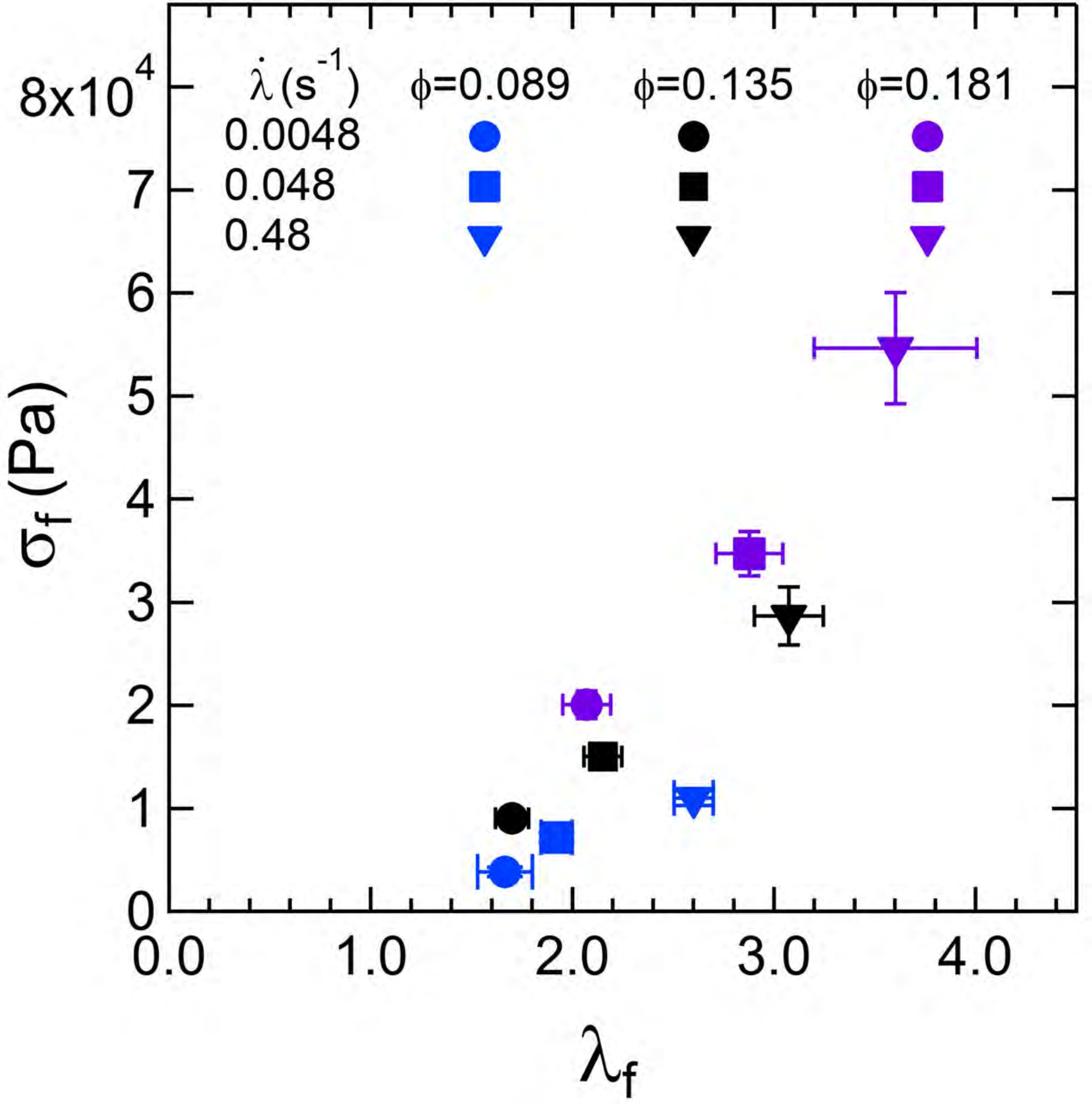}
		%\vspace{-3em}
		\caption{ Fracture stress ($\sigma_f$) as a function of maximum stretchability ($\lambda_f$) for $\phi$=0.089, 0.135, and 0.181 obtained from tensile experiments using stretch rate, $\dot{\lambda}$= 0.0048, 0.048 and 0.48~s$^{-1}$, respectively. }
		\label{PicSupConcTensileSfLf}
\end{center}
\end{figure}

\section{Estimation of Energy Release Rate}
The energy release rate ($\Gamma$) can be calculated based on the framework provided by Griffith. Based on his definition, the $\Gamma$ is defined as a change in the energy stored in material $W$ upon increasing the surface area.

\begin{equation}
\Gamma_0=\lim\limits_{dA\rightarrow 0}\left[\frac {U(r_{out})-U(r_{out}+dr_{out})}{dA}\right]
\label{Eqmain}
\end{equation}
Here, $U$ is the strain energy and $A$ is the surface area of cavity.
\subsection{WS approach}

The strain energy density function ($W(\lambda)$) of the neo-Hookean material is given as:
\begin{equation}
W(\lambda_1,\lambda_2,\lambda_3)=\frac{E}{6}\left(\lambda_1^2+\lambda_2^2+\lambda_3^2-3\right)
\label{EqNeoHookean}
\end{equation}
Due to large $L/r_{out}\approx$8.25, we can assume that the longitudinal stress (in 2 or $r$-direction) will be significantly lower than the hoop stress (in 1 or $L$-direction), resulting in the expansion of cavity in radial direction rather than the axial direction. Note that, we have observed the similar behavior in our experiments and is also reported previously for acrylic gels.\cite{Barney2019} Using the incompressibility condition, this  leads to the $\lambda_1=\lambda$, $\lambda_2=1$, and $\lambda_3=1/\lambda$. Plugging into the Eq.~\ref{EqNeoHookean}, we get:
\begin{equation}
W(\lambda)=\frac{E}{6}\left(\lambda^2+\frac{1}{\lambda^2}-2\right)
\label{EqCylinderStrainE}
\end{equation}

\begin{figure}[H]
	\begin{center}
		\includegraphics[width=4.5in]{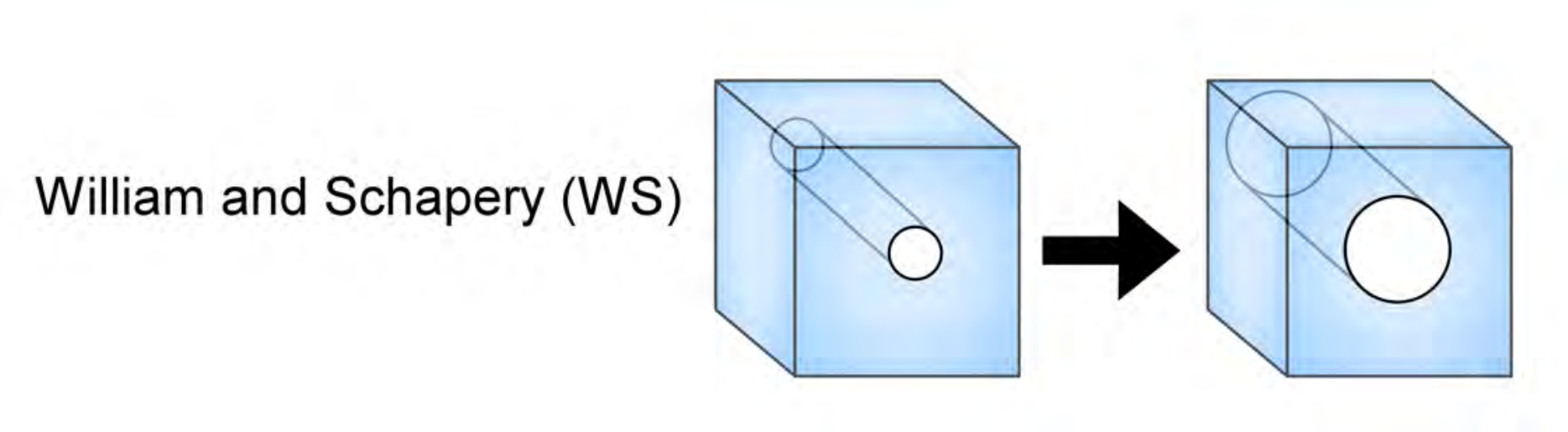}
		%\vspace{-3em}
		\caption{Schematic of cylindrical cavity expansion based on the framework provided by William and Schapery.\cite{Williams1965}  }
		\label{PicSupWS}
	\end{center}
\end{figure}
The WS approach is based on a cylindrical cavity expanding to a higher radius with the length remains unchanged. The initial cylindrical cavity has the length $L$ and radius $r_{out}$, which provides surface area as $2\pi r_{out}L$ and volume as $\pi r_{out}^2 L$. Upon application of a pressure ($P$), the energy stored in the system will be in the form of elastic strain energy leading to a change in the radius by $dr_{out}$. Note that, here we are assuming negligible change in the $L$ during cavity expansion due to high aspect ratio, as maintained in our experiments. Using Eq.~\ref{EqCylinderStrainE} for strain energy density, the change in strain energy of cylindrical cavity upon changing the cavity volume $dV$ can be estimated as:

\begin{equation}
dU=W(\lambda)dV=\frac{E}{6}\left(\lambda^2+\frac{1}{\lambda^2} -2\right) \pi r_{out} L dr_{out}
\end{equation}

Consequentiality, the change in the area $dA$ is $2 \pi  Ldr_{out}$, which estimates the energy release rate from Eq.~\ref{Eqmain} as:

\begin{equation}
\Gamma_0= \frac{Er_{out}}{12} \left(\lambda^2+\frac{1}{\lambda^2}-2\right)
\label{Eq10}
\end{equation}

In the low strain limit, $\lambda \rightarrow 1$, the $\Gamma_0$ approaches to $27 P^2 r_{out}/(4E)$.

\subsection{ Linear Elastic Mechanics approach}
According to the linear elastic mechanics, the small strain energy release rate can be estimated as, $\Gamma_0=(3\pi/4)(P_c^2r_{out}/ E)$.\cite{anderson2017fracture}

\begin{figure}[H]
	\begin{center}
		\includegraphics[width=3.5in]{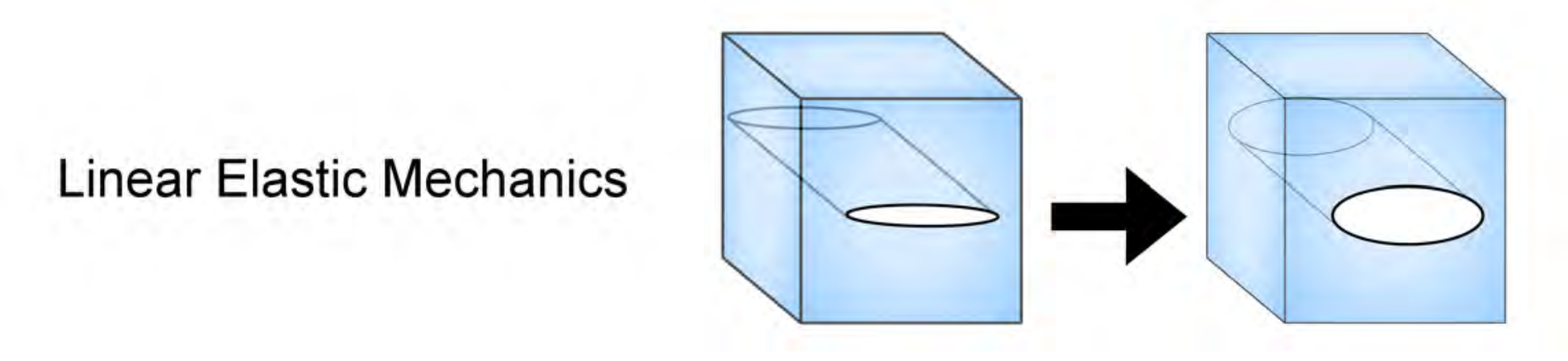}
		%\vspace{-3em}
		\caption{ Schematic of cylindrical cavity expansion based on the framework provided by linear elastic mechanics.\cite{anderson2017fracture} }
		\label{PicSupLEFM}
	\end{center}
\end{figure}

\begin{table}[H]
	\begin{center}
		\caption{Critical energy release rate estimated from linear elastic mechanics ($\Gamma_{0,LE}$) and Willam and Schapery approach ($\Gamma_{0,WS}$)  using cavitation experiments.   } 
		\label{TableConcCavitationFracture}
		\begin{tabular}{lll}
			\hline
			\multicolumn{1}{p{1.5cm}}{\centering $\phi$}  &  
			\multicolumn{1}{p{2.5cm}}{\centering $\Gamma_{0,LE}$ (J/m$^2$)} &
			\multicolumn{1}{p{2.5cm}}{\centering $\Gamma_{0,WS}$ (J/m$^2$)} \\

			\hline
			
			\multicolumn{1}{p{1.5cm}}{\centering 0.044} & \multicolumn{1}{p{2.5cm}}{\centering 3.17 $\pm$ 0.00} & 
			\multicolumn{1}{p{2.5cm}}{\centering 9.10 $\pm$ 0.10}\\

			\multicolumn{1}{p{1.5cm}}{\centering 0.067} & \multicolumn{1}{p{2.5cm}}{\centering 4.47 $\pm$ 0.00} & 
			\multicolumn{1}{p{2.5cm}}{\centering 12.80 $\pm$ 0.08}\\

			\multicolumn{1}{p{1.5cm}}{\centering 0.089} & \multicolumn{1}{p{2.5cm}}{\centering 5.67 $\pm$ 0.00} & 
			\multicolumn{1}{p{2.5cm}}{\centering 16.24 $\pm$ 0.15} \\

			\hline
		\end{tabular}
	\end{center}
\end{table}

\bibliography{Mishra_el_al_biblography}